# The Internet of People (IoP): A New Wave in Pervasive Mobile Computing

**Marco Conti\*, Andrea Passarella\*, Sajal K. Das\*\***

**\*** Italian National Research Council, IIT Institute, Pisa, Italy
**\*\*** Missouri University of Science and Technology, Computer Science Dept., Rolla MO, USA

**Abstract**: Cyber-Physical convergence, the fast expansion of the Internet at its edge, and tighter interactions between human users and their personal mobile devices push towards an *Internet* where the *human user becomes more central than ever*, and where their personal devices become their proxies in the cyber world, in addition to acting as a fundamental tool to sense the physical world. The *current Internet paradigm*, which is infrastructure-centric, *is not the right one* to cope with such emerging scenario with a wider range of applications. This calls for a *radically new* Internet paradigm, that we name the *Internet of People* (IoP), where the humans and their personal devices are not seen merely as end users of applications, but become active elements of the Internet. Note that IoP is not a replacement of the current Internet infrastructure, but it exploits legacy Internet services as (reliable) primitives to achieve end-to-end connectivity on a global-scale. In this visionary paper, we first discuss the key features of the IoP paradigm along with the underlying research issues and challenges. Then we present emerging networking and computing paradigms that are anticipating IoP.

## 1. Introduction

We are entering into an era where the boundaries between the physical and cyber worlds are more and more blurred, to the point that the two worlds become almost completely overlapped. In [CD2011] this phenomenon has been termed *Cyber-Physical Convergence*, as visually represented in Figure 1.

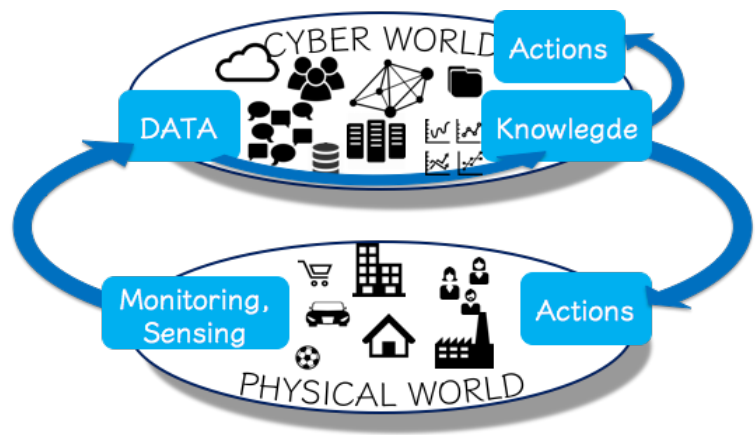

Figure 1: Cyber-Physical Convergence

The availability of massive amounts of data describing the processes, phenomena and behaviors in the physical world is a key element of Cyber-Physical Convergence. Data sensing the physical world are moved to the cyber world, where they are analyzed to generate appropriate representations of the corresponding physical entities and phenomena. This is done by extracting knowledge out of the massive amounts of data, typically through Big Data analytics techniques. Such knowledge is exploited by services and applications "living" in the cyber world. Based on this knowledge, services not only interact with each other, but also control the status of the physical world itself. In the first case, the effect of such interactions are typically modifications of the status of the cyber services, while in the latter case the effect is to modify or control (through actuators) the operations and performance of the physical system to drive them in the desired state(s). The vast availability of personal devices (e.g., smartphones) carried by humans, and of IoT devices (e.g., sensors) embedded in almost every physical object, together with the pervasiveness of wireless communication technologies, are fueling this unprecedented phenomenon.

Indeed, the Cyber-Physical Convergence is emerging as a game changer in the area of Pervasive and Mobile Computing [CD12b], for networks, services and applications oriented both to the domain of *human social interactions* ([CP17], [FPQSS16]) and to the domain of the *interaction between users and physical,* possibly critical*, infrastructures* [RLSS10].

The possibility to effectively and efficiently control and optimize the physical infrastructures from the cyber-world has received significant attention in the last ten years. As illustrated in Figure 2, through the pervasive use of sensors spread in the physical environment [GP15], the cyber layer continuously monitors the physical world. Sensed data are then transferred (through the Internet) to the cyber layer where, from the monitored data, knowledge is extracted to determine the "optimal" configuration and operating region of the physical infrastructures. Finally, through actuators, the cyber layer controls the physical infrastructures to make them perform as desired. The systems resulting from the convergence and close coupling of these two worlds are typically referred to as *Cyber Physical Systems* (CPS).

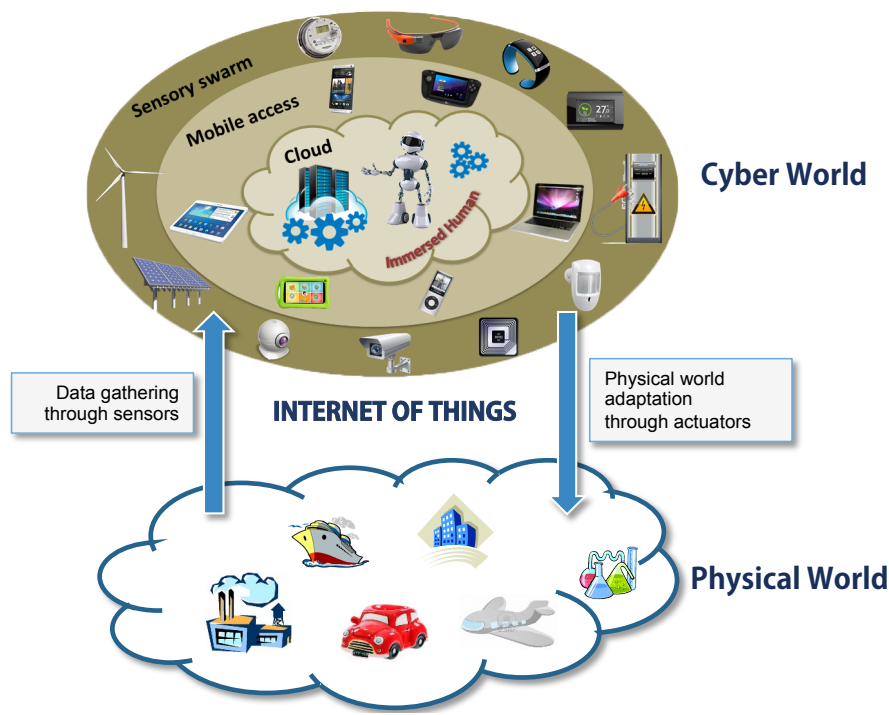

Figure 2: Cyber-Physical System (CPS)

Important CPS examples and applications are emerging in the everyday life, such as smart cities, smart transportation and driverless cars, Industry 4.0 and smart grids, smart healthcare, smart environments, and so on ([IKM10], [BR2012], [C16], [CD05], [CD07] [JGMP2014], [DAC2016], [LBK2015], [NSF]). In a smart city scenario, the objective of the cyber layer is to continuously monitor the status of the physical-city infrastructures to tailor them to the needs and improved experience of the citizens, in order to combine the efficiency in the usage of the resources with the required quality of the services; for example, by adjusting the traffic lights depending on the presence of cyclists [AFSAK2016], or the public transport system, according to the citizens' needs and preferences [ZDWZ2016]. Using a metaphor, we can say that the cyber layer is the *central nervous system* in a cyber-physical system and personal devices, such as the smartphones, are the (communication) doors between the physical and cyber layers. Through the pervasive use of information and communication technologies (e.g., sensors and actuators, IoT, big data and data analytics, cloud technologies, cyber-security, etc.), the cyber layer builds a "virtual" and interactive representation of the physical system, and use such representation to optimize and control the physical world ([ABC2013], [ABC2014], [EM15], [FGNAM2015], [HC2010], [KLKG15], [MKBD2012], [PTLG16], [TWBTGP2015], [RSB2016], [SH16], [SKEHBC2016], [JGMP14]). Clearly the Internet of Things (IoT) ([AIM2010], [B2014], [BGLLP2016], [GBMP2013], [MSDC2012], [WTPJ2016]) plays a vital role in the *Cyber-Physical Convergence* by guaranteeing a secure and energy-efficient transfer of information from the physical world to the cyber world, and vice-versa ([ABC2012], [AC2016], [CPAMM2016], [CRB2016], [DMP2015], [DWVT14], [XHL14], [RC2015], [SKEHBC2016]).

It is worth noting that, the role of human users in CPS is increasingly gaining attention in the sense that CPS are becoming complex socio-technical systems, and the human factor is becoming crucial in understanding and characterizing the behavior of CPS. In other words, humans are not only the final (passive) users of the CPS services but can also have an active role, similar to the ubiquitous presence and involvement of drivers in the transportation system, the workers in an industry, or the citizens of a city.

Along the above lines, another type of Cyber Physical Systems is emerging, which we hereafter denote as *Cyber-Physical Social Systems* (CPSS). CPSS are based on the fact that humans are social beings and their social relationships highly impact their behavior and actions, and thus, as depicted in Figure 3, the human social structures in the physical world (e.g., human relations network) and in the cyber world (e.g., online social networks) are also converging.

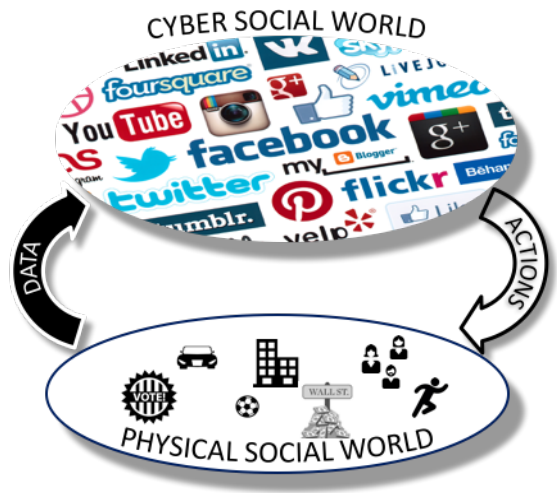

Figure 3: Cyber-Physical Social Systems (CPSS)

In CPSS, this convergence is primarily a side effect of social *interactions* happening in the physical world (e.g., through face-to-face interactions, physical meetings, and through spending time together), and social interactions happening online, e.g., over Online Social Networking sites. It is nowadays a common experience, also backed up by solid scientific results [SNM11, WDSGB11, CMJ11, DD17], that these two types of interactions intermingle with each other in a very interesting way. Meeting people in the physical world is significantly correlated with establishing social relationships in the cyber world [SNM11], particularly if the meeting happens in physical locations that are characterized by very specific features, which naturally define a social community of interest [BMNP13]. On the other hand, as shown in [CMJ11], movements of people outside of the confined area where they spend most of their time (e.g., "home location") are significantly correlated with the "home locations" of their social relationships in the Online Social Networks. In other words, when someone moves outside of their typical home location, they preferentially visit the home locations of their OSN friends. Other examples of convergent social behaviors between the physical and cyber social words are professional, business and financial relationships, voting and public opinion influence, organizing leisure time, etc.

There exist key common aspects between CPS and CPSS. The first common feature is that, in both cases, the convergence between the cyber and physical worlds happens by and large because of the special role played by the human users. In CPSS, this is evident due to its specific focus on human social relationships. As far as CPS are concerned, the extraction of knowledge out of data coming from the physical world, the way to interpret those data, the control decisions taken on how to modify the status of physical infrastructures, the way physical infrastructures are configured, heavily depend on the specific type of interaction between them and human users. For example, the same physical object may be configured in totally different ways depending on the specific users interacting with the object; different functions and operations may be enabled based on the trust the physical system is configured to have on the specific users. Similarly, physical infrastructures would be configured in different ways based on the way specific users access their services. In some cases, the CPS and CPSS concepts may overlap significantly. For example, this would be the case of Social Internet of Things systems (extending the basic concept originally proposed in [AIMN12]), where the interaction between physical objects happens by taking into consideration the social relationships between the owners or current users of the objects themselves. For example, a set of personal IoT devices owned by two users (e.g., two smart watches, or a smart watch and a TV set) would interact and share different information, allow or prevent mutual configurations, integrate more or less tightly into a unique smart environment, depending on the nature of social relationship (e.g., more or less tight) between the two users.

The second common feature is that in both CPS and CPSS, the users' personal devices act as proxies of their human users in the cyber world, and they are thus the main "bridges" between the two worlds. Many of the social interactions between the users in the cyber-world happen via personal devices that humans use to access OSN services. Moreover, the same personal devices can be used to monitor the users' individual and social activities (e.g., through opportunistic and participatory sensing, see Section 4.2) and thus provide to the social networking services in the cyber-world the knowledge for configuring themselves according to the user behaviors. In CPS, the interaction between physical objects and humans is again mediated by users' personal devices, which - again through pervasive sensing technologies - can also be used to monitor the users' behavior, and configure physical objects and infrastructures accordingly.

Therefore, Cyber Physical Convergence will be mainly human-centric, and all systems and technologies, designed with this vision in mind, must take the human individual and social behavior as one of the key design elements.

## 1.2 Internet of People: the New Internet paradigm for Cyber-Physical Convergence

We believe that a *new* Internet paradigm, that we hereafter denote as the *Internet of People (IoP)*, will be the key network substrate of Cyber-Physical Convergence of the future. The key IoP concept is captured in Figure 4.

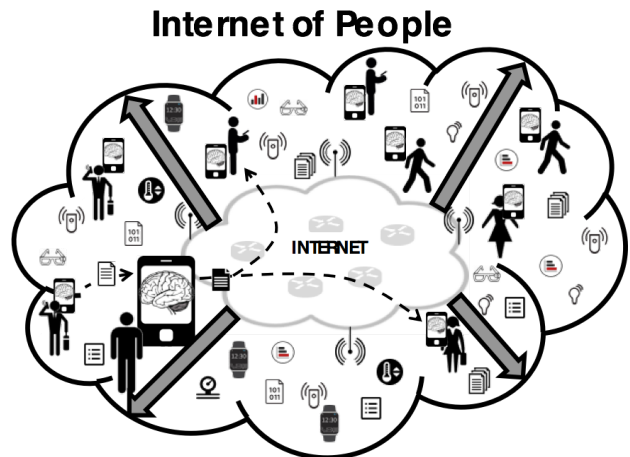

Figure 4. The Internet of People

In summary, Figure 4 depicts a networking environment where human users have plenty of connectivity opportunities either through the core Internet, with other users in proximity through self-organizing networking or with physical devices through IoT technologies (or combinations thereof). Even more importantly than mere connectivity, users are immersed in a huge amount of data that they could in principle access at any point in time. Data may come from remote users and devices, but most of the time data will have a strong locality dimension, and will enable interaction between users nearby, and between users and "local" things. The key novelty of this environment is the importance of users' personal devices, IoT devices embedded in physical objects, and the data they generate. This ensemble results in the exponential expansion of the Internet at the edge, and the need of *radically novel* networking primitives, which, as detailed in Section 3.1, cannot be realized through the conventional paradigms according to which the Internet core has been designed and evolved so far. The role of the IoP is to provide the necessary networking primitives through which users can (i) navigate this huge amount of data and access what is really needed, and (ii) exchange data and information with other users nearby based on their specific needs and actions (which may change dynamically over time). In all cases, IoP must embed trustworthiness in data access and in personal interactions. Note that the IoP is not prescriptive in terms of concrete applications. This is a common concept of both IoP and the legacy Internet, which has not been designed with a limited set of applications in mind. Rather, the IoP will naturally support the network and data access needs of the emerging Cyber-Physical Convergence vision.

More in detail, the IoP concept represented in Figure 4 is based on several observations. First, it is based on the observation that the *edge* of the Internet is expanding and evolving at a much higher pace than its core [C17, V14, M13]. The widespread diffusion of edge devices is indeed one of the enablers of the Cyber-Physical Convergence, as it is through these edge devices, which include both personal mobile devices of users and IoT devices embedded in physical objects, that data flows between the two worlds. This expansion *increases the "gravity" of the edge of the network* with respect to the core, and calls for novel networking paradigms where edge-centric networking assumes an increasingly important role, as opposed to the purely infrastructure-centric vision of the current Internet paradigm.

Second, it is based on the observation that, because the edge of the Internet is primarily populated by human personal devices, or by devices embedded in physical objects with which humans constantly interact, the Internet network paradigms (particularly at the edge of the network) need to take into consideration the behavior of the human users in the design of all networking functions, and the conventional, infrastructure-oriented paradigms used to design the core of the Internet are not always appropriate for the edge of the network. Both observations are discussed in more detail in Section 2.2.

As a key effect of these emerging trends, humans are more and more placed at the center of the technical systems they use and have an active role in their operations. Humans and the Internet devices, through which they communicate, become actors of a complex socio-technical ecosystem. One of the most intriguing effects of this convergence is that the *human becomes the center of the Internet system* and, for this reason, in [R2009], this paradigm change has been termed as an "Anti-Copernican Revolution". Looking forward, it is possible to foresee tighter levels of integration between humans and devices. For example, for some tasks, humans – and not devices – might be a better "resource" to be used in a cyber-physical convergence system. This is also a concept at the core of the Social Computer paradigm [RG13], where computers and humans are seen as "participant machinery" towards the realization of a given process.

Given the above emerging trends, humans and their devices (which behave as human proxies in the cyber world) become active, rather than passive, elements in the future Internet and for this reason we say that we are moving toward a world of *Internet of People (IoP)*. While the Internet core will continue to be one of the fundamental technologies, the IoP will not be the simple extension of the Internet coverage at the edge of the network. We strongly believe *IoP will be a new networking paradigm altogether*.

There exist several reasons, primarily related to the network design, why the *paradigms according to which the Internet networking primitives have been designed and evolved so far cannot cope well with the challenges brought by the environment described in Figure 4*. We discuss them in detail in Section 3.2. In addition, it is worth noting that putting humans, and their personal devices, at the center of the Internet makes it possible to address some of the major problems that severely affect the use of today's Internet *from the end user's perspective*. This can naturally be done by exploiting the strategies developed by the humans when they operate in the physical world, for conceiving the protocols used by their personal devices when they operate in the cyber world. For example, currently, a major concern is related to human control on personal data, the concentration of (personal) data in a few proprietary platforms and the related privacy protection issues. In IoP, by assigning an active role to the humans, is it possible to give them a higher control on their own data and prevent the abuse of personal data. To this end, in IoP solutions like Databox [HHC15] can be implemented as native network components, to create personal data spaces whereby the user can control access of third-party services to personal data. Moreover, in the current Internet it is difficult to establish the value and trust of the information we can collect from the network. Solutions clearly exist, but they have been included as "ex-post" patches, and are not native functions of the Internet networking primitives. This problem can be addressed in IoP by exploiting the social relationships existing among the (human) users. In the physical world of social relationships, trust is typically linked to the social closeness among people. Therefore, exploiting, inside the network, the models of human social relationship and trust, we can conceive effective policies for establishing trust between personal devices, with physical objects owned by specific users, and filtering unwanted/spam contents. Finally, the active role of humans inside the network may also help to increase the networks freedom reducing the risks of censorship. Interesting results in the area of censorship-resistant networking[1] can be inherited in the IoP vision.

This manuscript aims to highlight the research issues and challenges associated with next generation Internet. Specifically, Section 2 describes the difference between the evolutionary paths of development of the Internet core and access, and the revolutionary evolution of Internet at the edges brought by the IoP paradigm. In Section 3, we introduce the key features of the IoP, while in Section 4 we present some paradigms that represent first steps toward IoP. Section 5 discusses the IoP research challenges. Finally, Section 6 concludes the paper with further thoughts.

---

[1] See, e.g., the series of FOCI USENIX workshops on Free and Open Communications on the Internet.

## 2. The Internet of the Future

In addition to presenting the key IoP concept, Figure 4 also describes how we foresee the *Internet of the future*. The legacy Internet, made up of what we define as the Internet infrastructure, will continue to be the core of the network guaranteeing the connectivity between devices on a worldwide scale. As discussed below in Section 2.1, due to the untractable complexity of radically changing such a large-scale system affecting the global economy, the innovation in the Internet core will follow an evolutionary path. On the other hand, ad discussed in Section 2.2, it is more difficult to predict from the emerging trends, what will happen in the future at the edges of the network, which is currently expanding at a much faster rate than the core. However, we expect that the innovation there will be much more turbulent and radical, possibly following a more disruptive, revolutionary path.

### 2.1. The legacy-Internet evolution

We envisage that, in the future, the Internet core will be an information highway, mainly based on fiber optics [CCFJK2011], which may also use quantum technologies (e.g., quantum algorithms, quantum computing and quantum cryptography [BEMRR2007], [NC10], [RP00]), to connect the users to large data-centers ([WSXH2014], [WSXQ2016], [ZA2013]) providing large-scale Internet services. Power consumption of the data centers is already a critical issue (to make businesses profitable), and it is expected to significantly increase in the future; hence energy efficient policies will have a major role in the data-center design. Policies include a better usage of the available energy resources, included renewable energies (e.g., [MNBL2016]), and a reduction of the data-center consumption by exploiting virtualization techniques ([GBFJ2012], [AA2016], [AAMTW2016]).

More in general, virtualization techniques, by increasing the flexibility in the use of the network physical resources, are an asset for the evolution of the Internet core ([C2009], [KWKJ2009], [MSG16]). Network virtualization enables multiple service providers to share the same physical infrastructure, and allows the physical network resources to be used in the form of virtual networks with the required service levels ([KAI2016], [XTZ2015]). Network virtualization also supports the development of realistic environments for software testing before moving to the production phase [ADM2016]. Recently, the Software Defined Networks (SDN) paradigm is gaining momentum as a flexible network architecture, which simplifies the management of complex networks that may require dynamic reconfigurations [NMN14]. The SDN separates the control and the data plane. The control plane is programmable and uses an open interface -- the most notable is OpenFlow [KABPP2008] supported by the Open Network Foundation [OP] -- to program the packet forwarding devices in the data plane [XGHQ2015]. In this way, the network administrator can optimize the services, for example, to increase the Quality of Experience (QoE) of the end-users [GBFPR2015] or the security requirements [BOBBG2016]. The SDN-OpenFlow paradigm has gained momentum due to its applicability in several business domains: from data centers to cellular/service providers ([SNC2015], [SSPGC2016], [MNOG2016]). One of the key advantages of SDN is the possibility to deploy generic hardware devices in the core (instead of customized routers), and then dynamically reconfigure the devices behavior via the software control plane. The same concept is brought even further in Network Function Virtualization (NFV) [MSG16]. In NFV, any network function (not only routing, but also firewalls, DNS, etc.) can be programmed in software and dynamically instantiated on generic hardware nodes. The entire stack of typically core network functions can thus rely on (i) generic hardware and (ii) specialized software that dynamically (re-)programs nodes according to the specific functions the network needs to provide.

Another trend in the core Internet evolution is that, in the future Internet, data will have a major role. The access and dissemination of various types of content (from web pages to online social media), independently of its physical location, will be one of the dominant usages of the Internet. This means that the users will be mainly interested in accessing vast amounts of information, irrespective of their physical location. This data-centric communication model is not efficiently supported by the host-centric communication model of the legacy Internet. This has created a research community that looks at designing the Internet of the Future as an entirely content-centric infrastructure, typically denoted as *Information Centric Networking* (ICN) [XVS14]. The general concept of ICN has already been instantiated in various architectures ([ZABJ2014], [KCC07], [TP12]), also running on resource-limited devices ([SKS14], [ABCM16]). Multiple protocols addressing a wide range of ICN issues are currently subject of investigation ([GGMP2013], [ORS2012], [RLSL2016], [YAAW2014], [ZLL2013], [ZZRL2015]).

The last aspect of evolution of the Internet core is related to access. Specifically, this is where the Internet core "meets" the edges of the Internet. In the Internet of the future, the access to the core will be dominated by the 5G wireless technologies to provide high-speed ubiquitous connectivity [N15]. In order to cope with the exponential increase of data demand [C17] and the requirements set for 5G wireless [N15], different approaches are foreseen. Specifically, one approach consists of continuing with the trend of infrastructure densification, and targeting a massive deployment of femto-cells ([ACD12], [EOD2016]). This implies increasing the density of cellular base stations, to cope with the density increase of mobile devices, i.e., bringing the cellular infrastructure closer and closer to each and every individual device. However, there exist concerns on whether we are hitting fundamental limits at the physical layer, beyond which densification and other PHY layer improvements (e.g., massive MIMO) will not be able to help [AZD16]. MmWave is expected to provide extreme improvements from this standpoint [RRE14]. However, the density of mmWave base stations will have to be at least comparable to that of femto-cells, thus falling in the same type of problems. In parallel, the development of device-to-device (D2D) communications [AQM14] aims to make a more efficient use of the spectrum for supporting Proximity-based services, ProSe ([3GPPP], [LAGR14]). In all these cases, centralized coordination of the access configuration may become cumbersome, and distributed algorithms where decisions are delegated to users' devices are explored [WCA16].

The use of heterogeneous networks (HetNets) in the access part of the network corresponds to the same infrastructure-centric vision [WCA16]. In this case, users' devices dynamically switch between the available access technologies (e.g., WiFi or cellular), based on their conditions, trying to offload the cellular spectrum [RDCP15] [BMZP15] [RVBCD16]. Stretching this approach, delayed offloading has also been proposed to relieve cellular networks from delay-tolerant traffic. Specifically, in delayed offloading, bulks of delay-tolerant traffic are exchanged with edge devices only when the latter can access a higher-capacity infrastructure with respect to cellular, typically a WiFi hotspot. These forms of delayed offloading are already embedded in all modern smartphones operating systems. Currently, there is no clear consensus as to the benefits of delayed offloading and this paradigm is under active investigations ([MS14], [VBP16], [KK16], [PCP17]). An alternative approach to the cellular network offloading is based on the exploitation of infrastructure-less networks, in particular, the opportunistic network paradigm [WLLC12]. In opportunistic networks, D2D communications (not necessarily controlled by infrastructure elements, but possibly self-organized by mobile nodes themselves) are exploited, for instance to avoid repeated downloads of popular content on mobile devices collocated in the same physical area. Instead, the content is downloaded on only a few seed nodes, which then disseminate it to other interested devices, possibly exploiting the natural mobility of human users, through D2D communications. All these approaches tackle possible shortages of cellular bandwidth either in the access or in the backhaul. It is to be noted that, while the prospects of cellular capacity might seem to guarantee practically unlimited bandwidth to all possible devices, it is unlikely that this will be the case in the medium term. Traffic analyses [C17] show that traffic demand is going to increase exponentially, while cellular capacity is going to increase only linearly over time. Therefore, it is very much possible that the bandwidth crunches similar to the well-known ones that have affected 3G networks, might occur also with 4G and 5G technologies [AT10].

In order to cope with possible bandwidth shortages, the researchers are also looking at low-cost short-range solutions based on the unlicensed and uncongested visible light spectrum. The emergence of this solution is connected with the rapid increase in the usage of light emission diodes (LEDs), which are able to switch, at a fast rate, among different light intensity levels. This enables the possibility of achieving very high data rates up to some Gbps [PFHM2015]. The widespread availability of the LED technology on the vehicles makes visible light communication (VLC) a promising technology for Internet of vehicles [BMZC2016].

Together with enormous bandwidth increase, one target of 5G technologies is to provide very short delays and small jitters, in order to support real-time Internet applications along with infrastructure-based technologies. Typical targets in terms of latency are to arrive at a round-trip delay of about 1 ms. This is the typical latency for tactile steering and control of (real and virtual) objects. Achieving this target, together with highly reliable and robust communication links, will open up the era of the *tactile Internet* which can revolutionize several sectors of the society, such as health care (e.g., support to physically impaired people), mobility (e.g., full automatic driving), robotics and manufacturing, smart grid, etc. [F2014].

## 2.2. The Internet Edges revolution

The trends of evolution at the edges of the Internet are less easy to predict. However, some key points are evident, which will play important roles in defining the type of (r)evolution the Internet edges will undergo. Specifically, as discussed in the following, the conventional infrastructure-centric approach used to define the core-Internet primitives is not appropriate anymore when one considers these points together.

First of all, it is interesting to observe that the Internet at the edge is expanding at a much faster rate than the core. Mobile devices already dominate the Internet access; the diffusion of personal (mobile) devices and pervasive networking and computing technologies is expected to exponentially increase in the next few years. For example, CISCO foresees an eightfold increase of mobile data traffic between 2015 and 2020, with a compound annual growth rate (CAGR) of 53% [CISCO16]. This, together with the widespread penetration of Internet of Things (IoT), will contribute to a fast expansion of the Internet edges ([AIM2010], [B2014], [BGLLP2016], [GBMP2013], [MSDC2012], [WTPJ2016]). The expansion of the Internet at the edges applies both to the size of the network, as well as to the networking and computing capability of the network devices [C17]. As far as the former goes, the number of mobile devices connected to the Internet already significantly outnumbered the fixed devices [C17], and forecasts predict, in the next decade, an exponentially increasing number of mobile devices at the edge [V14, M13]. As far as the latter is concerned, mobile personal devices have plenty of resources to monitor and understand the context and behavior of their users in the physical world, perform complex analytical tasks on local data, communicate and exchange data with each other. In perspective, this tells us that the users will be immersed in a cyber-physical environment populated by a huge number of personal mobile devices and physical things which will generate, elaborate and transform a massive amount of data, outside the Internet core.

A complementary trend coupled with the expansion of the Internet at the edge is the *migration of network functionalities towards the edge* ([FLR2013], [MC2016]). The conventional Internet was (and is still) designed under the assumption that all networking functions are provided by the core network devices, which are responsible for "building" the required network abstractions. More recently, to cope with the increased complexity and performance requirements of the Internet applications, middleboxes and overlay network infrastructures (e.g., Content Delivery Networks or CDNs) became commonplace, to the point that we can consider them as an additional set of standard Internet functions. But, still, basically no such systems incorporate edge devices as key elements in the provision of Internet functions. Nevertheless in the Internet of the future (and, indeed, also today) the users' devices at the edge of the network will be extremely powerful and would be able to create their own local networking environments on-demand, without necessarily (or exclusively) relying on the Internet functionalities provided by core infrastructures. Therefore, the Internet of the future needs to embrace (mobile) edge devices as "first-class network nodes", and support design paradigms whereby the control over the deployment and operations of network functions may also be allocated to the users' devices, which autonomously decide how to cooperate with the core network to obtain networking services they need at a certain point in time. This trend is so evident that, as mentioned in Section 2.1, the use of the edge devices in implementing some network functions is also under development in the cellular technology with D2D communications services.

Last but not least, the Internet edge is going to be by and large human-centric. Most of the devices at the edge of the Internet are going to be either personal mobile devices, or IoT devices with which humans will directly interact to use physical devices and infrastructures. The tremendous diffusion of mobile personal devices, which are both data consumers and producers, are thus changing the way we interact with the Internet and the cyber world. Our personal mobile devices become our *proxies* in the cyber world. Since we carry them (almost) all the time with us throughout our daily activities, our personal devices can automatically detect the surrounding context including our activities, and self-configure accordingly ([BBH10], [PZC14], [ACD14]). Furthermore, because they are extremely sophisticated computing and networking devices, they can exchange data with each other, analyze the gathered data, and interact with other devices in the cyber world *on our behalf* [DAC2016]. Moreover, mobile devices will have sufficient resources to perform complex analytical tasks on local data (e.g., Big Data analytics and cognitive algorithms), thus effectively replicating, in the cyber world, the way their users would analyze and manage those data in the physical world.

All these trends make the conventional Internet paradigms, according to which its core is designed, not any longer appropriate nor sufficient to cope with the networking environment at the edge of the Internet. We thus need a radical paradigm shift, what we call the *Internet of People*, described in the following section.

## 3. Internet of People

Given the above observations on the evolution of the Internet, in this section we discuss in detail why the legacy (and current) Internet networking paradigms are not sufficient and the new Internet of People is needed (Section 3.1), and we discuss the key features we envision for IoP (Section 3.2).

### 3.1 The need for an IoP

As mentioned, the Internet has been conceived as a network infrastructure for reliable and fault-tolerant communications across a relatively limited number of sites [B64]. It has evolved with (i) an *infrastructure-centric* perspective, and (ii) the target to provide connectivity between any devices at a global scale. All the aspects related to the way users – humans in particular – use the Internet are not part of the fundamental Internet networking paradigms and have been taken into account outside the network infrastructure.

This *infrastructure-centric paradigm,* which *deals only with connectivity* aspects, presents *fundamental shortcomings* in the emerging cyber-physical convergence world. As discussed above, the Internet is expanding exponentially at its edges (much more than in its core), due to the ever-growing number of networked personal devices through which users "access" the cyber world. This makes the human user more at *the center* of the Internet. It requires a radical change of the Internet paradigm, which has been conceived and has evolved as an infrastructure-centric, rather than a human-centric, network. Furthermore, due to their tight bond with human users, in the Internet of People paradigm, personal mobile devices may act as their users would do when communicating, managing data, or computing. Indeed, IoP is device-centric, as users' devices take an active role in the network algorithms, as today's core nodes are active elements of the Internet algorithms. In IoP, users' devices decide the best enabling underlying technology, including the local networking with nearby devices (in addition to global networking through the current Internet), by exploiting self-organizing networking paradigms. A purely infrastructure-centric Internet does not take full advantage of spontaneous self-organizing forms of networking, which are capable of exploiting direct communications between (mobile) devices in physical proximity. By following human-centric communication patterns, self-organizing networking, rather than infrastructure-centric networking, is a more natural communication paradigm for the edge of the network. Already now, using exclusively infrastructure-centric networking leads to absurd situations, such as the need to have Internet connectivity and involve a handful of cloud services spread around the globe, to exchange a file –(through email or cloud storage services) between a couple of mobile devices in the same room. Or, to the need of delegating the configuration of local networking between nearby devices to possibly remote infrastructure elements, such as in long term evolution (LTE) D2D operations ([3GPPP], [LAGR14]). As the edge of the network is expanding so fast, a native support for these forms of spontaneous networking would be important.

*Data are also central in the IoP paradigm*, as data are the way through which information flows between the cyber and physical worlds, thereby enabling actions in both. Generally, the humans are the final destination of the data that personal devices may collect from the environment. The central role of data in this environment, and its huge amount, requires methods to filter data irrelevant for the humans and to assess the level of trust of the data, and of the users/devices from which data is received. In the current Internet paradigm, these aspects are not included in the network primitives, and are delegated to custom "add-ons" that each user is free to adopt or not.

In addition to collecting information from the cyber world around us, our smartphones that move along with us and act in the physical world, leave in the virtual world digital traces of our behavior such as: the human mobility patterns (e.g., our movements in the city), our social relations (e.g., the people we meet), our opinions, our consumption patterns, our economic/financial behaviors, and so on ([CCDD13], [CFT2016], [DDMG2015], [OVSBA2016]). The big data that encode our digital traces have a huge value from the commercial, social and scientific standpoints ([FSSST2016], [MCG2016], [PHWS2015], [PPSG15], [SZOT2016], [VNNR2015]). They can be used by decision makers to provide better services to citizens but, at the same time, if not managed in an appropriate way, may compromise our privacy ([BMCG14], [HDMP15], [REC15]). Therefore, privacy and security, and more generally the cyber-security, are essential elements in designing the Internet of the future ([WLWK2015], [BHAJ2016], [BR2015], [CRB2016], [LZD11a], [LZD11b]). Also in this case, conventional approaches to guarantee trust, privacy and security of information need to be rethought, and human-centric approaches are going to emerge from this standpoint. For example, techniques such as Databox [HHC15] are likely to become commonplace. In Databox, users delegate to their personal Databox system the control on access to their personal data. When Internet services require users' data, they negotiate with the personal Databoxes, and obtain access according to the policy specified by the users. This turns upside-down the paradigm according to which ownership of data is managed in today's Internet, where the (big) service providers collect personal data of users, and become de-facto owners of the personal data.

### 3.2 IoP features

The current Internet paradigm is neither data-, human-, device-centric, nor it supports self-organizing forms of networking, and therefore needs to be drastically changed in the cyber-physical convergence world. This calls for the new Internet of People (IoP) paradigm. In IoP, the humans and their personal devices are not only seen as final users of services and applications provided by the Internet service providers or operated by third parties, but will also be *active elements* of applications, services, and *network functions* provisioning. Furthermore, personal devices operate in the cyber world as proxies of humans and behave as humans do in the physical world. In IoP, personal devices are, therefore, key "nodes" of the network taking decision on the networking and data dissemination functions by possibly exploiting the model of the human behavior in the physical world. In many cases, network services will be device-centric, in the sense that users' devices will "take the initiative" and determine how the network needs to be configured and operated to satisfy their requirements.

They will dynamically configure the network to access devices nearby, as well as remote devices, thus having an active role in deciding how local networking resources as well as resources accessible through the core infrastructure will be used. Users' devices and core infrastructures will thus collaborate (often under devices' initiative) to provide the network functions.

Stretching this vision further, IoP embraces even a tighter integration between the Internet and humans, allowing humans themselves to contribute resources to the Internet functions. In IoP humans can also be "used" as network nodes, when their role is the most suitable one for the realization of specific parts of the IoP algorithms (i.e., the way IoP primitives are realized on the IoP nodes). This is an evolution towards the social-computer vision [RG13], where the human user is perceived as another entity of the computing and communication ecosystem, whose behavior can be modelled and predicted (clearly, up to a certain extent), and whose resources can be shared and exploited to optimize the operations of the system. As a starting rudimentary example of this vision, we may think of crowdsourcing systems, where humans are used to solve complex problems in a synergic way together with computers ([MZY14], [MYCH10], [KO14], [KTK15], [GWYW15], [RZZS2015], [RCBK2015]).

To summarize, the key characteristics of the IoP paradigm are:

- *IoP is human-centric,* and, as a consequence, is *multi-disciplinary*, as IoP algorithms should be based on quantitative models of the human individual and social behavior derived and validated in various research communities, such as sociology, anthropology, cognitive psychology, micro-economics, physics of complex systems, etc.

- *IoP is device-centric*, as users' devices are seen as "core IoP nodes", which are proxies of the humans in the cyber world, and host a significant part of the logic of the IoP algorithms.

- *IoP is data- and computing-oriented,* as IoP will naturally include primitives dealing with data management and data-centric computing, because data access, rather than connection to specific devices, is what humans will use the Internet for, most of the time.

- *IoP is also self-organizing,* as in IoP users' devices can establish spontaneous, infrastructure-less, networks with nearby devices, if local communications are the most effective ways to achieve a given task, e.g., exchanging data with the devices of other people that share the same location.

It is worth noting that *IoP is not a replacement of the current Internet.* The current Internet will remain the most suitable means of global-scale end-to-end connectivity, and IoP algorithms will use  it as one of the enabling technologies, but will develop a radically new human-centric and device-centric networking paradigm, which will exploit both local networking through self-organizing technologies, as well as global-scale networking through Internet technologies.

To better clarify our view, in the following we present a couple of examples to illustrate the IoP paradigm.

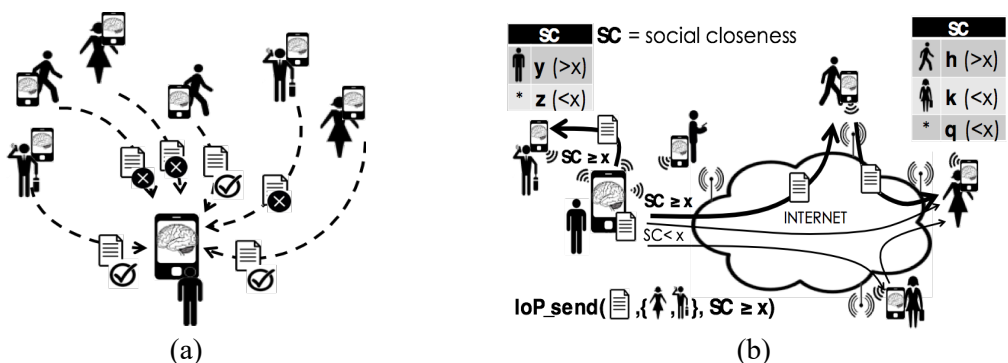

(a)                                          (b)

Figure 5. Examples of IoP human-centric filtering primitives

Specifically, Figure 5 presents two possible examples of IoP networking primitives, both related to data filtering. The first one (a) is a data-management primitive, in a self-organizing network, conceived from the standpoint of receivers; while the second one (b) is a connectivity primitive conceived from the standpoint of senders. Note that all the exemplified operations are carried out *automatically* by users' *devices*, acting as *proxies* of their *human users*. This requires that human behavioral models are embedded in the IoP algorithms. Through the primitive of case (a), a node fetches from local nearby nodes only data considered relevant for the local human user (i.e., data marked "√" with in the figure). To assess relevance, the node may use reference models of the human cognitive process for relevance assessment, e.g., cognitive heuristics [GHP11], as described in Section 4.3. In example (b), for data-filtering purposes, the receivers set a minimum value of *social closeness* (SC), *x*, with the nodes from which they *accept* data. Each receiver has a table in which it associates, for any possible sender, its social closeness. Nodes estimate SC by embedding reference models of the structure of human social relationships, for example, the use of ego-network models [ZSHD05, HD03] for information diffusion in online social networks [DAC15], as discussed in Section 4.4. The primitive "**IoP_send**(*data*, {set of receivers}, SC $\geq x$)" identifies IoP paths between the sender and the receivers, such that the *last* IoP node is "sufficiently close" to the receiver. In Figure 5(b), one destination can be reached at a sufficient SC through local self-organizing networking. The other one has to be reached through an intermediate IoP node. This node "validates" the data before forwarding it to the final destination (e.g., through SC between the sender and the intermediate node), such that its SC to the receiver is "transferred" to the data. In this example, the IoP nodes communicate over conventional multi-hop Internet paths, with state-of-the-art encryption, if needed. Thus, the example also illustrates the role of the current-Internet primitives in IoP.

## 4. First Steps towards IoP

While the current Internet is neither data-, human-, device-centric, nor it supports self-organizing forms of networking at the edge of the network, in the scientific literature we are observing the emergence of networking and computing paradigms which are precursors to IoP.

### 4.1 People-centric networking: self-organizing networks

In IoP, our smartphones via their wireless interfaces, such as LTE, 5G, WiFi, Bluetooth, or LR-WPANs ([ARS16], [BBBK2016], [DBA16], [DACDN11], [BCG02], [GRMM10]), connect us not only to a cyber world far away from us (e.g., Google data centers) but also to the cyber world around us. The latter is made up of the devices of the other people that are nearby -- our friends, our acquaintances, or just people who share the physical space around us. With them we might be interested in sharing information that has a special value for the space-time context in which it is exchanged. We may think, for example, of tourists visiting a city (see Figure 6) who can establish a dynamic social network, through their devices, to share contents (e.g., photos but also movie critics / impressions about the tour of the city) with nearby users in real time by exploiting the physical interactions of their personal mobile devices [YCC16]. Such dynamic social networks are referred to also as *Mobile Social Networks* (MSN) as they depend on the mobility patterns of people in the physical world, which bring groups of people (and hence their personal devices) in the same physical locations ([ACD14], [BMD13], [YL16], [DAC2016], [XLJD16]). It is interesting to observe that a location-based homophily exists among people that have social relationships [PK2016], i.e., people belonging a social community tend to visit the same locations and hence lean to be part also of the same dynamic social networks.

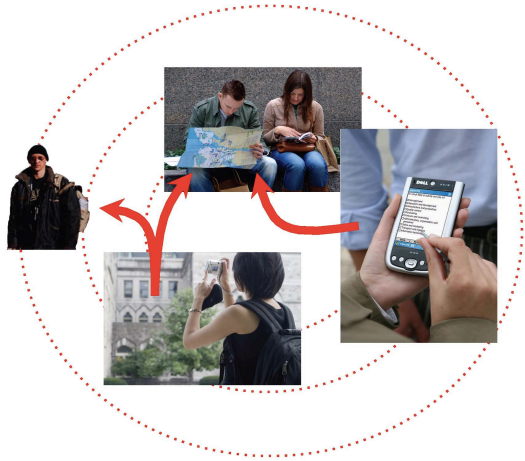

Figure 6: Mobile Social Networks

Our smartphones allow us not only to collect information around us but also carry and forward information to other people we meet [PPC2006] through various forms of self-organizing networking. In self-organizing networks with mobile nodes, the network is formed dynamically by the mobile nodes themselves, which automatically "create" multi-hop paths by exploiting movements of nodes. Therefore, it is through the movement of people that the information circulates in the cyber world. Hence, it is easy to recognize that multi-hop ad hoc wireless network paradigms, which have been extensively investigated in the last 15-20 years ([CBKM2015], [BCGS2013]), provide solid bases for IoP self-organizing networking. Specifically, starting from the mobile ad hoc networking (MANET) paradigm, multi-hop ad hoc wireless networking evolved in several paradigms, such as mesh, opportunistic, vehicular, and sensor networks, that now are building blocks for creating self-organizing networks. It can be expected that, in IoP, the various ad hoc paradigms will be mixed together and/or integrated with infrastructure-based networks, thus generating novel networking paradigms merging the self-* features properties of multi-hop ad hoc networks with the reliability and robustness features of infrastructure based networks [CG2013].

Among the various paradigms for self-organizing networks, opportunistic networking seems the most suitable one for IoP. Opportunistic networks are evolutions of the legacy MANET paradigm [CG13], conceived to cope better with dynamic network conditions and mobility of human users [PPC2006]. Indeed, while in MANETs an end-to-end path must exist between source-destination pairs, in opportunistic networks the source-destination paths are not required and data are transferred through the store-carry-and-forward paradigm exploiting the node's mobility. According to this paradigm, data are "moved" across the network not only through message passing amongst nodes, but also through the movements of the nodes themselves, which carry messages while waiting to enter in reciprocal radio range with either a message destination or a node suitable to reach the destination ([PPC2006], [TP13], [WHKD17]). Opportunistic networking has been traditionally considered as a mean to provide communication services mainly to portable devices (e.g., smartphones) in the absence of a fixed infrastructure (e.g., after a natural disaster [ABBB15]), or for supporting services, which depend on the (space and time) context, for which direct device-to-device communications are more effective [DAC2016]. In an opportunistic network, direct, physical contacts between nodes are opportunistically exploited to recognize and disseminate relevant information toward potentially interested nodes, without the need for centralized infrastructures or pre-computed paths from source to destination ([LDS03], [SP13], [SPR08], [VB00], [LD13], [SSKD13], [SDL15]). Often, in opportunistic networks, the nodes are personal devices and hence opportunistic networking is human-centric and device-centric, as the IoP (e.g., [BCP08], [DH09], [HCY11]). Specifically, human behavior has a major impact on the design of opportunistic-network protocols. First of all, the mobility of nodes corresponds to the mobility of their human owners, and thus the opportunistic-network paradigm highly depends on human mobility. Therefore, extensive studies have been carried out to characterize the properties of the human mobility and to define mobility models to design and evaluate the opportunistic-network protocols, e.g., [CHCD06] [KBCP11] [BP10] [PC13], [HSPH09], [KMS14], [LHKRC12], [RDCG14]. With reference to the IoP, an important research direction is related to investigating the relationship between human-mobility and social relationships in the physical world. Specifically, there is increasing evidence about the fact that mobility patterns are strongly influenced by social relationships between the users. For example, it has been found that people having a social relationship "attract each other", and thus mobility and social relationships are significantly correlated with each other [KBCP11, CMJ11]. Therefore, there is a growing interest in connecting the networking research with human-science research on human social network, which is a very relevant element for IoP.

Opportunistic networking, as IoP, is also data-centric. Indeed, opportunistic networks mimic, in the cyber world, the way information disseminates in the physical world through the contacts among people. To achieve this, personal devices should contribute part of their resources (e.g., energy, computing and storage capabilities) to the dissemination process ([BBPC14], [BCP10], [HFHC16], [LBSD15], [TPHH14], [WWXL16], [ZZWC13], [XLDWD16], [HCFDD13]). Thus, the data selection process must be very lightweight and able to perform a sharp distinction between data items, since only a very limited portion of them can be stored. The decision must be taken, based only on local observations, trying to balance local and global aspects, such as local memory availability and interest in the available data, the popularity of the data, the number of copies already replicated in nearby (and far away) nodes, etc. [BP13]. In other words, the devices must be able to adapt to the context by implementing autonomic policies [KRVS15]. In most proposals (e.g., [BCP10]), this is achieved through utility functions that rank the data items to store, taking into consideration local and global aspects, so as to balance effective delivery vs. resource consumption. Alternatively, the data items may be related to a certain geographical area within which their spreading is bounded [OK12]. In certain cases, opportunistic networking can also exploit infrastructure elements to help an effective dissemination process. This is the case of opportunistic offloading protocols [RDCP15] whereby the diffusion of content among groups of mobile devices primarily occurs via opportunistic contacts, but cellular coverage is exploited to deliver content to the users within a maximum deadline, if they have not received it via opportunistic contacts before. While these approaches have been conceived up to now from the standpoint of the cellular operators, in the IoP they could be re-conceived with a device-centric perspective.

## 4.2 People-centric sensing and computing

In the previous section, we have discussed the role of personal devices, including smartphones, in people-centric self-organizing networking. This section will discuss the evolution towards people-centric computing activities, which is driven by the increasing diffusion of smartphones that are full-fledged, communication, sensing and computing devices [DAC16]. The people-centric sensing (also known as *crowdsensing*) paradigm represents the first step in this evolution [SAS12]. According to this paradigm, which combines wireless communications and sensor networks with human daily life activities, people with their smart devices willingly or unwillingly represent potential sensing devices distributed across the physical space ([BSSD17], [BGSD17], [CNKLGHD17], [DDA11] [FD14], [LDPX16], [LKDT16], [KID14] [TSHDW17], [CELM06], [BEHP06], [E10], [LEMM08], [KID15], [SCH16]). A mobile phone, though not built specifically for sensing, can in fact readily function as sophisticated sensors by exploiting the camera (as video and image sensors), the microphone (as an acoustic sensor), the embedded GPS receivers (to sense location information), etc. Other embedded sensors such as gyroscopes, accelerometers and proximity sensors can collectively be used to estimate useful contextual information (e.g., whether the user is walking or traveling on a bicycle). In this way, the physical world can be sensed, without deploying a sensor network, by exploiting the billions of users' mobile devices/phones as location-aware data collection instruments for real world observations ([FRNDD12], [HCCL13] [MZY14]). For example, the microphones of smartphones can be used to provide a very detailed map of the noise pollution in a reference area [RCBK2015]. People-centric sensing is a human-centric paradigm for sensing the physical world: while humans move in the physical world, their devices sense the physical world and transfer the sensed information to the cyber world, where this information is used to build a virtual representation of the physical word that can be used to provide better services to the citizens.

Smartphones are the reference devices for people-centric sensing, however, this paradigm can be extended to include other forms of people-centric sensing, such as vehicular sensing. Vehicular sensing exploits the large set of sensors embedded in vehicles, the personal smartphones of the driver and passengers, and also ad-hoc sensors (e.g., for environmental monitoring ([HFZC16]) installed on vehicles to collect data at urban scale ([CCDD13], [EGHN08], [ZLYX11], [ZMLL15]).

People-centric sensing well exemplifies the cyber-physical convergence: while people really move into the physical environments, their devices sense the physical world and transfer the sensed information to the cyber world. It is clear that people-centric sensing perfectly fits into the IoP vision. In fact, the sensing devices move with people and hence the sensing of the physical world is tightly coupled with the human behavior and human social relationships ([RZZS2015], [RDP16]). While people-centric sensing provides, in principle, a huge sensing environment available to support context-aware human-centric services and applications, there are clearly huge challenges related to privacy of the data and the knowledge about the users' behavior that can be inferred. Privacy is, therefore, a major research challenge ([KD15], [LYYW15], [LZDT11], [PDDA14], [RLLWD14], [RRFCGDP15]) in this framework.

People-centric sensing is an example of a service that can be implemented by exploiting the personal-device resources. In [CK10] this paradigm is extended by defining the opportunistic computing paradigm to exploit and coordinate all personal-device resources, as well as other resources available in the physical environment, in order to provide localized computing and communication services that are tightly coupled with people and their devices. For example, a device may offer as a service its high-speed Internet connection to other devices nearby. Similarly, the possibility to offload a computational task from a resource-limited device to a nearby resource-rich device is also a research direction [FGP2016] [MZAD16] [SAZN12].

The opportunistic computing approach also envisages the possibility to create enhanced services by composing the services currently available on nearby devices [UKPC14], e.g., to deliver in an efficient way a large file to the cloud, a small IoT device can compose the compression service offered by a nearby device with the Internet access service provided by another device. This idea is refined and implemented in the SCAMPI architecture [PKOC12]. This architecture abstracts resources (computational power of personal devices, sensors, resources embedded in the local environment, etc.) as service components, and leverages the human social behavior, to opportunistically find, access and compose the services available in the surrounding environment to dynamically create a rich set of services.

Similar to the opportunistic computing, other paradigms, like fog computing [BMZA12] and cloudlet architectures [SBCD09] are pushing the intelligence towards the edge of the network by exploiting gateways, at the boundary between the access network and the Internet, with a goal to provide services to the users. However, opportunistic computing represents the first paradigm that is pushing directly the intelligence to the personal devices. Specifically, the mobile smart devices, by pooling their resources can start offering services as a sort of *mobile clouds* bringing services and resources closer to where they are needed [LMED15], thereby avoiding the bandwidth costs associated with accessing the services in the cloud [BBCM16], [CMP15]. In certain cases, this is also a way to save energy of mobile devices, as it is known that, for significant classes of applications (e.g., those requiring scalar computing) the energy cost of computing is typically much lower than the energy cost of wireless communications. However, for vector data computing such as video and image data sensing, the energy cost of computing may be comparable as well as more than energy consumption due to communications [LLD06], [LLD09], [LLLD06], [SLD12].

Coordination of local mobile devices forming a mobile cloud has been recently proposed in FemtoCloud [HAHZ15], where the controller (residing in a gateway) is entirely responsible to schedule parallel tasks execution on mobile devices.

The services provided by mobile clouds are mainly data-centric. The (personal) devices forming the mobile clouds manage most of the data generated locally, and provide data-centric services to each other based on such data. For example, in ([VPC16], [VPC17]) a scenario is considered where each device does not necessarily shares locally generated data (e.g., data sensed from the environment) with the other devices (or with a remote cloud service). Instead, each device uses distributed machine learning techniques to extract a partial model from its own data, which it then exchanges with other devices. Refined models are obtained by combining the partial models received from other devices.

Also in this case, it is possible to exploit hybrid approaches, where computation may be done both on the mobile devices nearby or on fixed infrastructure elements. Initial results in this direction have been presented in [CMP15]. The key challenge is how to design such schemes in a distributed way, under the control of the users' (mobile) devices.

## 4.3 Human-centric data collection

In the cyber-physical converged world, the active user participation in the process of data creation and diffusion creates a huge quantity of pervasive information stored in the personal devices around us. We are immersed in an invisible cloud of digital information, and while we interact, in the physical world, with the people close to us, in the cyber world our smartphones may interact with the personal devices of nearby people for sharing and collecting information on our behalf. It is worth noticing that, contrary to the Internet approach, it may not be always necessary to bring all these data to the centralized cloud storage facilities, where they are analyzed and accessed. Indeed, a considerable portion of this data will also be very contextualized, i.e., relevant only at specific times and/or geographic areas, and of interest only for specific groups of users. Therefore, it is more appropriate to conceive data management schemes where decisions on how to store, replicate and disseminate data is delegated to the mobile devices in proximity of where data are generated and, most of the time, consumed.

In such a context, personal devices act as the *avatars* of their respective users. They allow their owners to explore congested cyber information landscapes by collecting the available information, filtering it, and presenting relevant data to the human brain of their owners. It is, in the end, the brain of the user the final recipient of the collected information. At personal devices, a key challenge is therefore how to select, in an effective way, the information to present to the users, avoiding to flood them with huge amount of (useless) information that the human brain cannot handle. In ([CMP13], [LPCP15]), a methodology is proposed to overcome this problem by directly embedding in the personal devices the cognitive processes used by the human brain to filter out irrelevant information. In other worlds, as represented in Figure 7, our personal devices must learn how to operate in the virtual world for selecting what is important for us, as our brain would do. Specifically, Figure 7 represents the paradigm shift resulting from using such a human-centric approach to the design of data dissemination protocols. Conceptually, in the current paradigm (represented in the top part of the figure) personal devices select information to be presented to the users based on their explicit requests and actions (such as configuration of the devices). In a radically novel paradigm (represented at the bottom of the figure), a user's personal device embeds human cognitive models, of how the user brain would process available data, and is thus able to automatically decide which data is relevant (and thus to be presented) to the user, out of the possibly huge set of data available in the network environment.

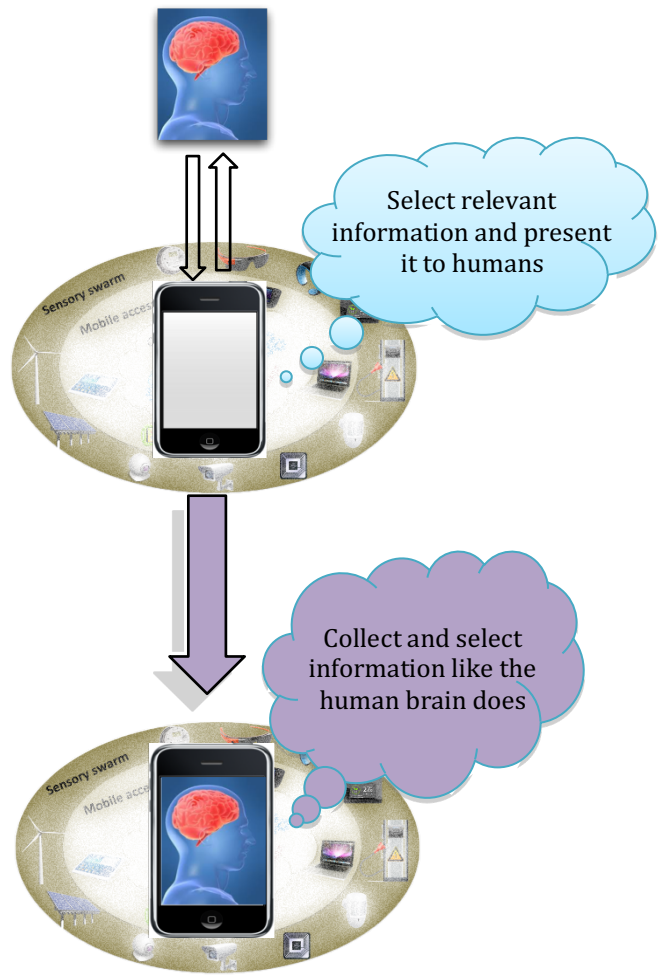

Figure 7: Information selection by exploiting the human brain algorithms

The human brain performs this task using so called cognitive heuristics, i.e., simple, rapid, yet very effective schemes to assess the relevance of information under partial knowledge ([GG92], [GG96]). Cognitive heuristics are fast, frugal and adaptive strategies of the brain that allow humans to face complex situations by addressing simpler problems [MGG10]. By exploiting the cognitive heuristics, the human brain is able to swiftly contextualize the stimuli it is subject to, identify the relevant features and knowledge to be considered, assert the relevance of perceived information, and finally select the most useful data, even when only partial information is available. Hence, despite their simplicity, cognitive heuristics are indispensable psychological tools that are very effective in solving decision-making problems like information selection and acquisition [MGG10].

The use of cognitive heuristics to develop effective algorithms for data collection in the cyber-physical world has been first proposed in [CMP13], where, two of the several cognitive models present in the cognitive psychology literature are considered [GG96] [G08] [MGSG10]: the *Recognition Heuristic* and the *Take the Best Heuristic.*

The *Recognition* heuristic assumes that merely *recognizing* an object is sufficient to determine its relevance [GG02], where an object is recognized if the user "sees" it a sufficient number of times[2]. Therefore, in [CMP13] the recognition-heuristic principle is

---

2    "Seeing" should be intended as a very general concept, which is not necessarily bound to visual perception. For example, the same heuristic can also be applied to concepts, e.g., which might be encountered by a person exploited to let each personal device rapidly decide the utility of taking one data item available in the cyberspace, instead of another. More precisely, data are stored in the personal devices of other users moving in the same physical space and a new data is available as soon as devices are in direct contact through their wireless interfaces. Building upon the recognition heuristic, an algorithm is proposed that is inspired by the *Take-the-Best* cognitive scheme. This algorithm uses the reference model, in the cognitive literature, of Goldstein and Gigerenzer [GG96] and exploits the recognition heuristic in order to simplify and limit the complexity of the data selection task. This is done by recursive creation of small subsets of all the discovered data from which the relevant data is sorted out. The bottom line idea is to assess relevance of data items based on the recognition heuristic, prioritize available data items accordingly, and store only the most relevant available data items until the local cache used for the dissemination process is full. The work presented in [CMP13] proves the suitability and effectiveness of these heuristics in problems, like data dissemination in cyber-physical world, where every node has only a partial knowledge about its environment. However, due to the large number of status information the nodes have to maintain, this approach may suffer from scalability problems. In [VPCP15] an improved solution addressed those scalability problems. Nodes exploit, beyond cognitive heuristics, (i) a local aggregate measure about items diffusion and (ii) a stochastic mechanism to choose which data items should be replicated. The resulting solution shows to be more efficient and scalable than the previous one and, thanks to the stochastic decision-making mechanism, independent of specific scenario configurations.

Generally, in characterizing the human interest it is assumed that the users' interest is rather static and quite simple to describe [CMP13]. Users are supposed to be interested in predefined content categories (e.g., sports, movies, etc.) and therefore their devices should collect all the contents related to those categories. However, in the reality the users' interests can change over time, as a result of a knowledge acquisition process that is also the effect of social interactions between them. To model the dynamicity of users' interests based on cognitive schemes, in [CMPR13] a semantic network is introduced to represent the data stored in the users' devices. A semantic network is a graph, where vertices are the semantic concepts (e.g., the tags associated to data items) and edges represent the connections that exist among them (e.g., the logical link between two tags). The semantic description could be the case, for instance, of tagged photos on Flickr and Instagram, or messages annotated with "hashtags" in Twitter and Facebook. Note that in [CMPR13] the semantic network represents concepts, while data items stored by the personal devices are not directly represented in the semantic network. Instead, they are associated to a set of concepts, which they refer to (i.e., concepts are considered as "tags", and data items can be tagged with multiple concepts). When personal devices "encounter" each other, they first exchange information extracted from the semantic network, and this drives the exchange of specific data items, as explained in the following. This process is the same behind exchange of knowledge and data associated to knowledge between human beings. Specifically, in [MVCP16] by exploiting the semantic-network representation, the selection of information to exchange starts from the concepts in common between the two nodes, similar to the way a real discussion between humans typically starts. Then, the semantic network of each personal device is navigated starting from these common concepts, and each node extracts a selection of concepts to pass over to the other node. Finally, the contents stored at the nodes, referring to the exchanged semantic concepts, are also exchanged between the nodes. Therefore, users' (dynamic) interests drive the dissemination of semantic data, which drives the dissemination of content. Specifically, in the proposed approach, the semantic data are disseminated among users' devices, based on the current interests of the users by exploiting a set of algorithms based on cognitive heuristics. The novel aspect of this work is therefore the use of cognitive heuristics to optimize the joint dissemination of semantic information and associated contents.

Cognitive heuristics have also been applied to provide cost effective algorithms for

---

while reading a book.

solving several other problems emerging in the cyber-physical world. Cognitive heuristics and models of dynamic memory structures have been used in [MPCA15] for crowdsourcing in the smart city environments. In [MPC15] the social cognitive heuristics have been applied to develop a decentralized solution that allows each user's mobile device to identify social relationships of its user, while in [MPC2015] the social circle heuristic [PRH05] is exploited to develop a dissemination policy for opportunistic networks where each node stores data items that are relevant for itself and for other nodes in its social context.

### 4.4 The ego-network model and data dissemination in IoP

As shown in the Example b) in Section 3.2, in the IoP, models of human relationships are the bases for developing data distribution primitives with trust guarantees. Specifically, according to the IoP approach, the design of effective primitives requires the availability of quantitative models of the human social relationships.

Models of the human social relationships have been extensively investigated in the social sciences literature. In [ACPD17], after reviewing one of the most promising quantitative models for characterizing the human social relationships, the authors investigate how this model can be applied to design and evaluate novel data dissemination protocols in IoP. Specifically, in [ACPD17], the authors focus on the ego-network model, which has been proposed in the anthropology literature, to describe the social relationships of an individual (ego) with its social peers (alters). Ego networks are one of the key concepts to study the microscopic properties of personal social networks [APCD15]. There exist different definitions of ego networks, corresponding to different approaches in analyzing them. Hereafter we used the definition proposed in [SDBA12]: an *ego network* is formed of a single individual (*ego*) and the other users directly connected to it (*alters*).

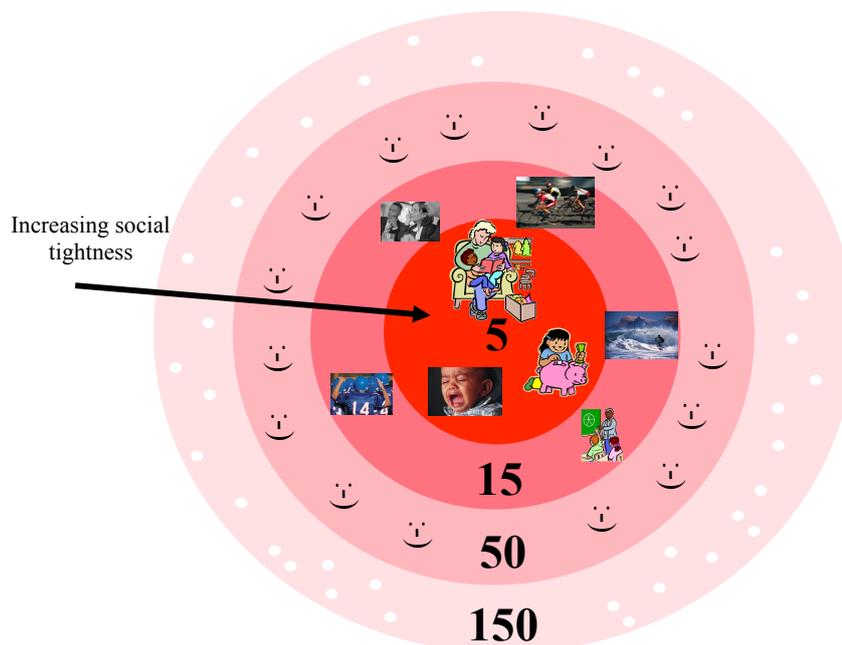

Figure 8: Dunbar ego-network schematic representation (from [ZSHD05]). Alters are represented inside concentric layers around the ego (at the center of the picture). Membership in layers is determined by the strength of the social tie with the ego.

According to the ego-network model, a fundamental cognitive constraint in the human's personal social network is the Dunbar's number [D93]. This is the number of relationships that an ego *actively* maintains in its network over time, i.e., the relationships for which a minimum frequency of interactions (typically, one per year) is maintained over time. The Dunbar's number in offline ego networks is known to be limited by the cognitive constraints of the human brain, and by the limited time that people can spend in socializing. In addition, it is known that cognitive constraints lead people to unevenly distribute the emotional intensity on their relationships. As depicted in Figure 8, this results in a hierarchical structure of inclusive 'social circles' of alters around the ego with characteristic size and level of tie strength (the strength of the social relationship). Specifically, in the reference ego-network model [ZSHD05], there is an inner circle (called *support clique*) of 5 alters on average, which are considered the best friends of the ego. These alters are contacted at least once a week, and are the people from whom the ego seeks help in case of emotional distress or financial disaster. Then, there is a second layer of 15 alters, called *sympathy group* (which includes the support clique) containing close friends of the ego, those contacted at least once a month. After this layer, we find a group of 50 alters, called *affinity group* or *band* that contains an extended group of friends. The last circle, called *active network*, contains on average 150 alters (the Dunbar's number) contacted at least once a year. These people represent the social relationships that the ego maintains actively, spending a non-negligible amount of time and cognitive resources interacting with them so as to prevent the corresponding social relationships decaying over time. The sizes of ego network circles form a typical pattern of 5–15–50–150 alters, with a scaling ratio between adjacent circles around 3. This pattern is considered one of the distinctive features of human social networks. Evidence to support the existence of Dunbar's number and the described ego-network structure in offline social networks has come from a number of ethnographic and sociological sources [D93], [ZSHD05]. More recently, results have also been shown on the presence of Dunbar's number and the ego network structure in phone-call networks [MMLM13], [GQRZ17] and in online social networks [APCD15], [DACP15]. In [ALPC16], the authors have investigated the impact of the ego network structures on information diffusion by exploiting the online social links. Specifically, they have shown that, by considering the structural properties of ego networks, it is possible to accurately model information diffusion both over individual social links, as well at the entire network level, i.e., it is possible to accurately model information "cascades" [GHF13]. Moreover, the authors have analyzed how trusted information diffuses in OSNs, assuming that the tie strength between the nodes (which, in turn, determines the structure of ego networks) is a good proxy to measure the reciprocal trust. Interestingly, they have shown that not using social links over a certain level of trust drastically limits the information spread; for example, when only very strong ties are used, the information diffuses up to only 3% of the nodes. However, inserting even a single social relationship per ego, at a level of trust below the threshold, can drastically increase information diffusion. This is consistent with the well-known Granovetter's results, that showed that strong ties can carry a significant amount of information, although weak ties are also important for acquiring diversity of information [G73]. Finally, when information diffusion is driven by trust, the average length of the shortest paths is more than twice the one obtained when all social links can be used for dissemination. Other analyses in the latter case have highlighted that in OSN also, users are separated by about 6 (or less) degrees of separation. The results presented in [ALPC16] show that when we need trustworthy "paths" to communicate in OSN, we are more than twice as far away from each other.

The relevance of ego-network structure in the study of OSN properties opens up several research directions, which are relevant for IoP. The dependence of information diffusion on the trust of social relationships can have a significant impact on the design of novel social networking platforms such as Distributed Online Social Networks (DOSN), as discussed in [ACPD17]. DOSN implements the functionalities of OSN platforms, but in a completely decentralized way. DOSN are human-centric and device-centric in order to maintain the control on personal data. In fact, personal data of the users and the content they exchange are stored directly on their devices, without the need of any third-party server to operate the social networking platform. This provides much more control on the user over their personal information, but requires caching and replication techniques to guarantee data availability [CSGP14] [DDGR16].

The structural properties of the ego-network can also be used to improve data availability in other types of social-oriented networking systems, such as Mobile Social Networks (MSN). Knowledge about the structural properties of ego networks could both improve the accuracy of data dissemination in MSN, and thus can make them more easily adaptable to different social contexts. Examples of this approach are data dissemination schemes based on the social circle heuristic, as described in Section 4.2.

## 5. IoP Research Challenges

Although the previous sections highlighted the existing literature with research ideas which anticipate the Internet of People concepts, several research challenges have still to be addressed to arrive at defining the full IoP paradigm: from the definition of an IoP framework and architecture to the human-centric design of its networking and computing primitives. Hereafter, we present and discuss a list of IoP open research challenges that are not exhaustive. As in the legacy Internet, we first start with the architectural issues and algorithms and protocols for implementing end-to-end communication services. Then we focus on novel IoP aspects such as in-network data and computing services and the overall management of resources of the IoP ecosystem. Finally, we discuss the challenges associated with the performance evaluation models and techniques that need to be developed and exploited to investigate a complex socio-technical system as the IoP.

### 5.1 IoP framework and architecture

The first building block in the IoP definition is the identification of IoP nodes and of their communication, data and computing services. The IoP nodes include, in addition to devices provided by the infrastructure operators (as in the conventional Internet), user devices, physical objects ("things"), and in a long-term view, also human users.[3] These devices should provide, in addition to typical communication services such as unicast, multicast, anycast communications, services to support data management (e.g., data replication, data dissemination, data advertisement), as well as computing (e.g., data analytics).

The second building block is the design of a network architecture suitable for IoP. This may need a drastically re-thinking of the Internet architecture, in particular, as the IoP is primarily designed for humans, and for "naturally" supporting their behavior in the physical world. This opens up several research questions. What is a correct architectural approach to embed human behavioral models in the IoP protocols? Is the current separation of functions across the Internet stack still meaningful for IoP? How do we modularize functions into layers and stacks, and how do we deal with cross-layer information and interactions?

### 5.2 IoP Network protocols

ICT enabling technologies (both wired and wireless) provide the basic primitives for communications to occur. As in the legacy Internet, by exploiting the enabling technologies, the network protocols should, first of all, provide end-to-end connectivity among IoP nodes to support the same type of services available in the current Internet: unicast, multicast or anycast. However, in IoP, network protocols, which are implemented also in the users' devices, should also decide which underlying enabling technology should be used. They will decide, for example, whether to use Internet connections, or local self-organizing networking between nearby devices, without any assistance of the Internet infrastructure. Therefore, the network protocols will also implement the IoP device-centric vision, as nodes will autonomously decide how to configure the network around them.

The algorithms and protocols for IoP networking and data exchange are not driven exclusively by the need to optimize network resource usage (as in the design of the legacy

---

[3] For example, see the social computer paradigm [RG13].

Internet systems). In the converged cyber-physical environment, the needs of the users (e.g., trusting the content received over an Internet path) may be even more important than pure network performance (e.g., optimizing the throughput of that specific path). Therefore, networking protocols should behave the way their human users would do if interacting with each other in the physical world. Furthermore, in IoP, networking services will be tightly coupled with humans, and hence appropriate models of the human behavior taken from relevant research communities must be embedded in the protocol design. This way, such models can become solid foundations for making the operations of the IoP protocols truly human-centric. Indeed, as already shown in the literature of self-organizing networks, exploiting the models, which characterize the human "social" mobility [KBCP11], is highly beneficial for networks formed by mobile users' personal devices. For example, to define forms of "group-cast" primitives, where groups represent mobile social networks (i.e., local, transient communities of users meeting in a certain place to perform together a certain social activity). A less investigated aspect is related to the use of human social models to define human-centric protocols for communication services that also include core infrastructure and are deployed at the level of the entire network. For example, in case a path involving nodes with a certain level of *social closeness* with each other (and thus trust) is needed, the algorithms will use models of social relationships between the IoP nodes to identify the suitable paths, and select the best one. This is the case exemplified in Figure 5(b).

As humans are social beings, we anticipate that relevant models for designing human-centric protocols could be found in domains dealing with the study of individuals (e.g., cognitive psychology [GHP11, GTA99, S05]), as well as social communities (e.g., sociology and anthropology [HD03], [ZSHD05], [RD11]). Furthermore, models derived in micro-economics [FG07, FF02, CF06] can be used to model resource availability and the way nodes can "trade" them, as well as the trust associated with them. In addition to human social models, large networks of individuals and devices interacting together can be modeled by exploiting models of complex system physics (e.g., [BA99, DM03, C07, DLGD17]). Such models represent, through compact mathematical indices, the important (or "centrality") of nodes (and thus, users) in the network. These indices can be used to design human-centric data forwarding protocols or data replication policies. Initial attempts in this direction have been explored in the framework of opportunistic networking [HCY11, DH09], where indices such as betweenness centrality of nodes are used to rank them, and decide which are the most appropriate nodes to use as forwarders for messages.

### 5.3 IoP in-network data and computing services

Data management will be one of the key functions that IoP nodes will have to support as a network service, the same way today's Internet nodes support routing and forwarding. As already highlighted in the previous sections, data filtering is a key service that IoP nodes should provide. However, other IoP primitives will deal with data management, e.g., for supporting access to data, irrespective of the location where it is generated/stored, or to guarantee data privacy. Therefore, primitives in this class will include dissemination of data to the interested users, replication of data to support its availability and access to it (possibly within specified time bounds), advertisement of data availability and search for data. Human-centric aspects of these primitives will define how humans access and filter data, as well as how they share, advertise and trust information exchanged between each other.

As explained in Section 4.3, cognitive models constitute a very promising research direction for building data-management services capable of filtering among a huge amount of data, those relevant for the final user, i.e., the human brain. This approach has the potential to become a core feature in the IoP design. However, the approach explored so far (cognitive heuristics) is only one of the many ways to model human cognitive functions. Sometimes such modeling approaches are even contradictory to each other (e.g., cognitive heuristics vs. Bayesian approaches [S05, SF17]), and therefore it is not yet clear how to build a comprehensive representation of the human cognitive processes, which could be used as a foundational component of IoP solutions.

We envisage that IoP nodes should also provide *computing* services on the available data as key native functions, i.e., IoP should natively provide both computing and networking primitives. This set of primitives will be conceived to make IoP a distributed computing system, starting from the data available on a heterogeneous set of IoP nodes. The human-centric dimension will define how computing resources are managed locally, as well as how they are shared between the IoP nodes. Initial results in this direction are opportunistic computing approaches, described in Section 4.2. It is worth noticing that this would extend the current areas related to fog and mobile edge computing. Specifically, in IoP - as in fog and mobile edge computing - mobile nodes may form distributed, dynamic mobile clouds used to perform generic computing tasks on demand. Infrastructure elements (similar to current fog gateways) may also be used. The paradigm shift would be to define device- and human-centric policies to organize these mobile clouds based on the specific user' needs. Thus, IoP concepts can also be clearly exploited to perform computation on data (e.g., extracting knowledge from raw data), but what is the right approach to embed such functions in IoP primitives?

## 5.4 Resources management and trust models
IoP will be built on a heterogeneous set of resources, provided by devices owned by specific users, devices not bound to any specific user, and human users themselves. What is the correct approach to represent these resources, advertise and orchestrate them, and make them available in the IoP ecosystem to build complex network functions? In IoP network resources may be very dynamic and heterogeneous, and only partly controlled by (more or less) trusted operators. On the other hand, it will be in the interest of all users that IoP is an ecosystem where own resources can be shared in a fair way, in return of better network services built thanks to these shared pools of resources. In this perspective, which models should IoP adopt to establish trust and facilitate cooperation between parties, such that users will be confident in sharing and using each other resources? To model resource availability and trade, and the trust associated with it, we can use models derived in micro-economics [FG07, FF02, CF06]. Interestingly, differently from conventional game-theoretical models, these approaches take into consideration the effect of social relationships in the way cooperation among humans develops. It has been shown that patterns of cooperation between individuals, which determine sharing of resources, are not correctly modelled by considering only the actors' self-interest (as in conventional game theoretical approaches) but need to be modelled taking also the social dimension into account. This literature is important for IoP, as these models will be used to describe the possibility to access shared resources from the standpoint of individual nodes. For example, if a resource is contributed by other users nearby or by devices of users with whom an ego has some type of social relationship, the possibility of exploiting that resource in IoP would be described through these models. Similarly, human-centric relationships could also be exploited to design effective incentive schemes for collaboration between users through their personal devices.

## 5.5 IoP Performance Evaluation
The use of the IoP paradigm creates a complex socio-technical system, where quantitative models of the human behavior are embedded in the protocols and algorithms. Performance models are therefore important tools to understand the behavior of the resulting system(s) made up not only of devices but also of humans. The complexity of the models calls for effective techniques that can tackle these challenges. Indeed, quite likely, the IoP systems will be too complex to be investigated with accurate (and solvable) analytical models. On the other hand, accurate simulation models could also be too complex, thus requiring a huge amount of computational resources (and hence a long simulation time) to investigate few units of simulated time. To address this challenge, in [BCMP13] the authors exploit the *Hierarchical Modelling* (HM) technique, a structured approach aimed at reducing the modelling complexity of highly structured socio-technical systems, such as Internet of People [CM98]. Specifically, HM -- starting from a model of the overall system (original model) -- uses a three-step procedure that can be applied recursively. In the first *decomposition step*, the model is partitioned into sub-models. Each sub-model is then analyzed in isolation to obtain a simpler representation of that sub-model (step 2), where the complexity of the simplified model depends on the type of study. The *aggregation step* is the last HM step (step 3): in the original model, the simplified models derived in step 2 replaces each sub-model; the resulting model (aggregate model) is then solved either analytically or via simulation. Several combinations of analysis and simulation can be applied in step 2 and step 3. In very complex systems, as that analyzed in [BCMP13], the Hybrid Simulation approach has been shown quite effective: each sub-model (typically a protocol) is analytically solved and then the overall aggregate system is solved via simulation. Specifically, in [BCMP13] the authors have used HM to investigate the properties of the protocol presented in [CMP13] for data dissemination in the cyberspace exploiting the recognition-heuristic. Specifically, the problem to investigate is the performance analysis of the cognitive-based data-dissemination protocol in a scenario where thousands (or more) of personal devices move in the cyberspace, following the movement of their users in the physical world (e.g., an entire country). The users belong to a multitude of different physical communities, where each community of users (e.g., a group of users that share strong social connections) moves inside a physical space, which is separated by the space of the other communities. In addition, a small subset of the users, called "travelers", move across different communities, thus bridging them and allowing the delivery of data from one group to another. For example, this scenario may represent citizens of different cities, where the citizens typically move only within their city geographical area and only a limited number of citizens (the travelers) leave this area to visit other cities (and then come back). As the first step, the authors have subdivided the model into sub-models, each corresponding to a community (a city); then, as the second step, the authors developed an effective analytical model for studying the data dissemination when all users of a community move only inside the physical space of the community [BCMP12]. Next (the third step) they developed a HM model – to study the data dissemination in the reference scenario – in which the movements of the travelers are explicitly modelled, while the data dissemination process inside a community is described with the steady-state performance indices obtained from the analytical model. In this way, in [BCMP13], the authors have been able to study a data dissemination process up to a country level with few millions of users involved in the dissemination process.

More in general, the study of opportunistic-network[4] performance has several similarities with IoP analysis (in both cases the human behavior is a key factor in understanding the system behavior) and therefore the basic results and methodologies developed in the performance studies of opportunistic networks represent a basis for IoP evaluation, as well. In particular, the performance of both systems is highly dependent on the human mobility patterns. In fact, the mobility of the users determines the dynamics of encounters between users' devices and hence the performance of the network protocols. Indeed, the encounters duration affects the amount of data that can be exchanged between two nodes, while their frequency and pattern affect delivery delay and delivery probability, respectively. For this reason, opportunistic-network researchers have devoted a lot of efforts to collect human mobility traces and to use them to understand the properties of the human mobility ([MIT], [CHCD06], [WDCA11], [GPR11], [HNS12], [PC13], [DDMG2015], [GTNF16]). Specifically, it is now well recognized that several parameters describing human mobility and encounters follow a power law distribution, such as the distance covered by a movement between two places; the visiting time, return time, and visiting frequency of a location; the contact duration and inter-contact time between pairs of persons (see, e.g., [KBCP11] for an extensive survey). Based on these results, several mobility models have been implemented ([HSPH09], [BP10], [LHKR12], [KMS14]), amongst others), that are able to supply synthetic traces satisfying the power law distribution and well approximating real traces, which can be used for experimentation and validation of solutions designed for opportunistic networks. Starting from these models, the researchers have developed several analytical and/or simulation models for investigating the performance of opportunistic network protocols. In the literature, we can find several analytical performance studies focusing on routing protocols and content dissemination algorithms for opportunistic networks, which are reference points for the performance analysis of IoP protocols ([ZNKT07], [BCP14], [PS15], [BCP15], [PK16]).

---

[4] Including all type of networks, e.g., vehicular networks, which use opportunistic encounters depending on the human behavior.

## 6. Conclusions

In this paper we have discussed the Internet evolution towards the Internet of People (IoP), a complex socio-technical system where humans, with their personal devices, are the key nodes of the network. Indeed, user devices become *proxies* of their users in the cyber world: they communicate, exchange and manage data on behalf of their users, and thus should behave the way their human users would do if interacting with each other in the physical world. Therefore, we argue that in IoP we need to take into consideration the human behavior as a *structural design paradigm*, rather than as an afterthought. In IoP, the human behavior is not only considered at the services and application layers but also, according to a user-centric design, it permeates the entire "network stack" (although the concept of protocol stack itself would need to be re-thought). This is because people are no longer mere users of network technologies and services, designed exclusively with engineering optimization parameters in mind, but their behavior becomes one of the key elements for designing all facets of the Internet technologies. Therefore, the algorithms and protocols for IoP networking and data exchange are not driven exclusively by the need to optimize network resource usage (as in the design of legacy Internet systems) but are quantitatively modelled and considered as a first-class design parameter in order to cater to the human needs.

Since various facets of the human behavior need to be "coded" into the networking protocols, we need to link our traditional technology-oriented perspective closely to human-centric sciences (describing human behavior) for designing IoP networking and data exchange mechanisms that are human-centric. Human behavior has to be embedded into the networking protocols and devices logic, to influence the operations of the human networking functions. Adopting such an inter-disciplinary approach is not an easy task, as it requires to bridge very different scientific disciplines. We believe that a cornerstone to fruitfully follow this approach is to seek *quantitative* mathematical models (rather than qualitative descriptive models) or algorithmic definitions to describe the needed facets of the human behavior. Using quantitative approaches is fundamental, as it provides an appropriate common "language" that strongly supports and facilitates the discussion among diverse research communities. Ideally, these models and algorithms should emerge from the "non-ICT" community studying that specific facet of the human behavior, as that research community can truly challenge and validate these models. Once validated, being expressed in a mathematical and/or algorithmic form, the models are amenable to be directly "embedded" into IoP protocols and systems. Using this methodology, the models of the human behavior are not interpretations given by the Internet researchers, but rather well established scientific descriptions, validated by the appropriate research communities. Specifically, the IoP network systems will be designed taking into account several interacting non-ICT dimensions, such as: social sciences, cognitive psychology, complex networks, and microeconomics.

*Social sciences* model the way users establish social relationships, how they trust each other, and how they are prepared to share resources with each other. *Cognitive psychology* describes, among others, how human beings perceive and interact with data, how they assess relevance of information, how they exchange it when interacting, and how they extract knowledge out of it. Data-centric Internet systems for mobile networks have already been proposed, where these models are exploited to efficiently guide information diffusion among users [MVCP16]. Very useful models have been derived in the area of *complex network analysis* [C07], describing, for example, human social relationships with compact graph descriptions, amenable to characterize human behavioral properties and exploit them in the design of networking solutions. Finally, microeconomics models how humans negotiate the use of infrastructure and content resources, trade and share them. While these models will be developed and validated with an interdisciplinary work led by non-ICT communities, ICT technologies, like Big Data Analytics and Deep Learning, will have a major role in validating and tuning these models in the IoP context [APCD15].

We would like to stress the fact that the proposed approach to the design of IoP is not yet another bio-inspired networking design wave. Due to the fact that user devices act as proxies of their users, and the human brain is often the final destination of the information collected in the Internet, embedding efficient models of human behavior in the core design of networking systems is a natural way to make devices behave as their human users would do if faced with the same choices and decisions. Moreover, this approach is not confined to designing human-centered applications. Instead, the inter-disciplinary approach of IoP impacts all conventional layers of the network stack above the enabling communication technologies, and brings advantage to all layers, as shown by the mentioned examples.

Due to the increasing importance of data in the Internet ecosystem, and in particular for the human users, IoP will be a data-centric network. *IoP embraces data-oriented Internet functions.* The IoP concepts can be used not only to re-design "conventional" networking functions, but also to design data-centric network functions as a core part of IoP, as already demonstrated in some initial cases described in the paper. IoP will naturally include primitives dealing with data management and data-centric computing, because data access, rather than connection to specific devices, is what humans will mostly use the Internet for. *IoP is also self-organizing,* as IoP user devices can establish spontaneous, infrastructure-less, networks with nearby devices, if local communications are the most effective way to achieve a given task, such as exchanging data with the devices of other people that share the same physical space.

While IoP is a radically new paradigm for the Internet, some systems exist which represent the first steps toward realization of IoP. In particular, a human-centric approach has been already used in an effective way in the recent past for the design of network protocols. Specifically, we have discussed crowdsensing, mobile social networking, cognitive-based data dissemination in the pervasive Internet and data diffusion in online social networks (OSNs). However, these systems represent only preliminary building blocks for the IoP framework, and need to be significantly refined and integrated. To achieve this, several research challenges need to be addressed. This paper identified a first set of open research challenges ranging from the IoP architecture to the design of human-centric and data-centric IoP algorithms and protocols including cross layer issues, such as IoP resource management and trust models. We consider these as initial elements of a more complex IoP environment. While, until now, they have been investigated in isolation, IoP provides a unifying concept for their harmonic development. In addition, in the view of the new concept of IoP, each of these elements will clearly need to be reconsidered, further investigated and developed.